# The Hubbard Model:
# A Computer Lab in the Strong Coupling Region

Donald M. Esterling[1]

VeritasCNC    Portland, OR 97209

**Abstract**

A challenge – and opportunity – is offered to the Hubbard Model community of solutions extant for the strong coupling region.  A rigorous and quantitatively demanding test –  a Computer Lab – is presented based on certain exact results for the strong coupling limit derived in a companion paper.  The test offers rich insights into the essential physics of the model, specifically the quasiparticle energy and width of the spectral weight function in the lower Hubbard band.  The width of the spectral weight function, related to electron scattering, is large and has a strong momentum dependence, even for infinite coupling. The test is applied to mean field solutions as well as some one dimensional results.  In each case, substantial problems are exposed.  The method is also used to determine a "best" (i.e. moment preserving) solution, subject to only one assumption of a single dominant peak for the lower Hubbard band spectral weight function.

**Introduction**

The (Fermi) Hubbard model [1, 2] is a deceptively simple statement of a many body problem – becoming the object of theoretical treatments over many decades (e.g. as reviewed in Esterling and Dubin [3], Cyrot [4]; Imada, et al. 1998 [5]; LeBlanc et al. [6]).  The model has been applied to various experimental phenomenon, but with conflicting conclusions about the underlying physics leading to "a great need for reasonably unbiased methods for determining the physics from a spectral analysis" (Gunnarsson et al. [7]).  Further, the single band Hubbard model (the most commonly treated case) is itself a simplification of materials with d-bands, p-band hybridization, electron-phonon effects, random alloy effects and, of course, inter-site coulomb repulsion.  An "unbiased" arbiter of proposed model solutions, clean without conflating  how well the model itself replicates the actual material complexity, would be of some value.

The Web of Science [8] citation index lists almost 20,000 articles on the topic "Hubbard model", many of which cite and/or introduce approximate solution methods. A singular goal here is to offer a rigorous, quantitatively demanding set of benchmark tests of these approximate solutions – a Computer Lab – specific to the important strong coupling region, that is valid across all dimensions and lattice structures and which captures some of the essential physics of the model through an analysis of the spectral weight function (SWF) of the model itself. The Computer Lab, while at times algebraically tedious, offers the virtue of simplicity in technique and mathematics.

---

[1] E-mail:  don@veritascnc.com

The analog simulations of strongly correlated material systems via ultra cold quantum gases in optical lattices [9] offers a complementary approach to this purely quantitative test.  The use of the analog method to probe the momentum and frequency dependent SWF in the strongly coupled region of the Hubbard model [10] would provide a useful bridge between the methods.

Solutions to the Hubbard model in the strong coupling region have been a challenge [3-6, 29].  While there are exact results for the one dimensional case [11], there are few if any exact benchmarks for general dimensions and lattice structures.  The genesis of the Computer Lab tests starts with moment relations for the LHB SWF derived by Harris and Lange [12].  The moments are expressed in terms of higher order equal time correlation functions.  In a companion paper [13], these higher order terms are exactly expressed as explicit functions of the single particle equal time correlation function.  The moments, in the infinite coupling limit, then depend only on the single particle Green's function.  The relations hold for any dimension and any lattice structure, but in the strong coupling and zero temperature limit.  The notations in this paper follows that of [13].

The exact method does not invoke "decoupling" of the higher order moments as in the past [14-16].  Koch, Sangiovanni and Gunnarsson [17] offer certain sum rules for quantum cluster theories, but the rules assume the integrity of cluster mean field theories.  The sum rules here (moments) make no assumptions about the underlying theory. Matteo and Claveau have recently offered an alternative moment solution [18]  but their resulting sum rule integrates out both frequency and momentum and assumes a simple two pole approximation to the SWF for finite U.

It is well known that moments of a function alone cannot, in general, uniquely determine the explicit function.  However exact moments of a function can provide insight into the quality of an approximation to that function.

The Computer Lab flows as follows. The single particle equal time correlation function in the strong coupling limit is determined from a proposed solution, then used to compute the various LHB SWF moments and the results compared with a direct computation of the LHB SWF moments.  The comparison strategy is outlined in Fig. 1 below.  If the solution is "true", the results should be similar if not the same, since the two methods follow from the same definition of the SWF and the exact (infinite coupling or U, zero temperature) expressions.  Deviations indicate numerical and possible qualitative errors in the solution.  While restricted to the strong coupling limit and zero temperature, sizeable deviations or errors will indicate problems into the strong coupling (finite coupling or U) region and finite temperatures.

Many solutions to the Hubbard model focus on intermediate coupling (U on the order of the hopping or $\Delta$, each defined in Eqn. (1) of [13]) near the Mott transition.  There are no quantitative guidelines in this region.  Solutions that claim to be valid in the intermediate region typically do so by asserting the solution is well behaved in the weak and strong coupling regions. The weak coupling region is easy enough to recover. This Computer Lab puts the solutions to the test in the strong coupling region.

If an explicit functional form is assumed, the moments can also be used to generate a self-consistent solution. In a later section, the LHB SWF is obtained with the *single assumption* of a single peak, specifically – though not necessarily – a Gaussian form. This "best" (i.e. LHB moment-preserving) SWF provides insights into any Hubbard model solution with a single peak in the lower (and upper) Hubbard bands. Both this Single Peak solution and a Computer Lab solution (for a mean field solution method [19-22]) yield a large width to the SWF as the electron density approaches one electron per site, when computed with the "Indirect method" of Fig. 1 below. This result and its implications are discussed below.

## Atomic Limit Redux

A companion paper [13] dissects the "atomic limit" term often loosely used by various researchers and objects to the oft-stated contention that the atomic limit can be solved "exactly" or is trivial. In fact the physically interesting atomic limits (**AL1** and **AL2** of [13]) where $k_BT$ (temperature) is much less than the hopping ($\Delta$ in Eqn. (1) of [13]) which is, in turn, much less than the intra-atomic coupling (U in Eqn. (1) of [13]) is non-trivial, **non-local** and likely never to be solved exactly [4,23]. The usual reference to "atomic limit" in fact is **AL0** of [13] which is the high temperature case $\Delta \ll k_bT$. This case has limited physical interest. For the remainder of this paper, unless noted otherwise, the term "atomic limit" will refer to either of the physically relevant limits of **AL1** or **AL2**. The important challenge is to find a viable solution at low or zero temperature.

To recall, **AL1** has $k_bT$ identically zero, $\Delta > 0$ (but $\to 0$) and U finite. So $\Delta \gg k_BT$. **AL2** has $k_BT$ identically zero, $\Delta > 0$ and U infinite. So again $\Delta \gg k_BT$. **AL2** is not usually considered as an "atomic limit", but it shares the property that $\Delta/U \to 0$ at zero temperature. Solutions that claim to be correct for $\Delta/U \to 0$ (either large U, small $\Delta$ or moderate U, vanishing $\Delta$) – that is, claim to have a physically correct (or exact) expansion of some quantity in $\Delta/U$, must be so for *both* **AL1** and **AL2**. The expansion should be equally correct if $\Delta$ is vanishingly small and U finite or if $\Delta$ is finite and U goes to infinity.

The first systematic expansion of the self energy in $\Delta/U$ was developed by Esterling and Lange [24]. Similar to the Green's function functional derivative expansion for weakly interacting systems [25], Esterling and Lange derived a functional derivative expansion of the self energy in powers of $\Delta/U$. Shortly thereafter, Esterling himself [23] pointed out the basic deficiency of such an expansion, generalizing the concern to any proposed solution that involves an expansion in $\Delta/U$ as summarized above, specifically that if the expansion is valid then the proposed solution must become exact in the **AL2** (infinite U, finite $\Delta$) case. This is a strong assertion and unlikely to be true.

The problem involves any time-dependent function. Time can enter as an inverse energy or frequency. Frequencies of interest are of order $\Delta$. Expansions in powers of $\Delta$ can invoke terms which go like

Δ /(frequency) which, in general, cannot be neglected.  Ignoring such apparently higher order terms in Δ can lead to contradictions [23].

The results derived in the companion paper [13] rely on the computation of *equal time* correlation functions.  In this case, exact results can be obtained. (More precisely, higher order equal time correlation functions can be exactly expressed as functions of the single particle correlation function).  In fact, as noted by Esterling [13], from a purely dimensional argument, the equal time correlation functions are the same in **AL1** or **AL2**.

On the other hand, any solution to the Hubbard model which asserts an expansion in inverse powers of U (e.g. [7, 26-30] )[2] must conform *exactly* to the results of this Computer Lab in the infinite U, finite Δ limit.  This is the challenge – and opportunity – offered by the Computer Lab test.

### Exact Results as a Computer Lab:  DMFT

The general strategy of using exact results for the moments of the LHB SWF in the infinite U, zero temperature limit is outlined in Figure 1.

The exact moment results can serve as a severe test of any proposed solution to the Hubbard model in the strong coupling region.  An often-cited solution method is the single site Dynamic Mean Field Theory (DMFT) ) [19-22] and its extension to clusters [31].  A critical assumption in DMFT is a local self-energy [31].  The cluster extensions accommodate a non-local self-energy, but still retain a cell-wise averaging over momentum space which is a similar assumption.  Here the focus is on DMFT but some discussion of the cluster solutions will follow in a later Section.

The results are for a 1D linear chain, a 2D square array and a 3D simple cubic lattice with nearest neighbor hopping, finite Δ, infinite U and unit lattice spacing.  The method and formulae are easily extended to higher dimensions and alterative lattice structures. Focus is on the momentum distribution and the SWF width, each computed as in [3]. The paramagnetic state for the electron density ($n_\sigma = n_{-\sigma} = n$) is assumed throughout unless explicitly noted otherwise.

The self-energy for the single site DMFT Hubbard model solution in the strong coupling limit has been indentified as the atomic limit self-energy [11]. This self-energy $\Sigma$ is independent of momentum k and, for infinite U, becomes

---

[2] This is a far from exhaustive list of perturbation solutions involving Δ/U.  Further, Metzner [26] notes that "Clearly a perturbation expansion terminated at some finite order is only applicable when the hopping matrix $t_{ij}$ is small compared to the temperature T, i.e. either for very narrow bands or for extremely high temperatures."  Unfortunately, others have invoked the Metzner "linked cluster expansion" without this caveat.  See, for example, Dai, Haule and Kotliar [30].

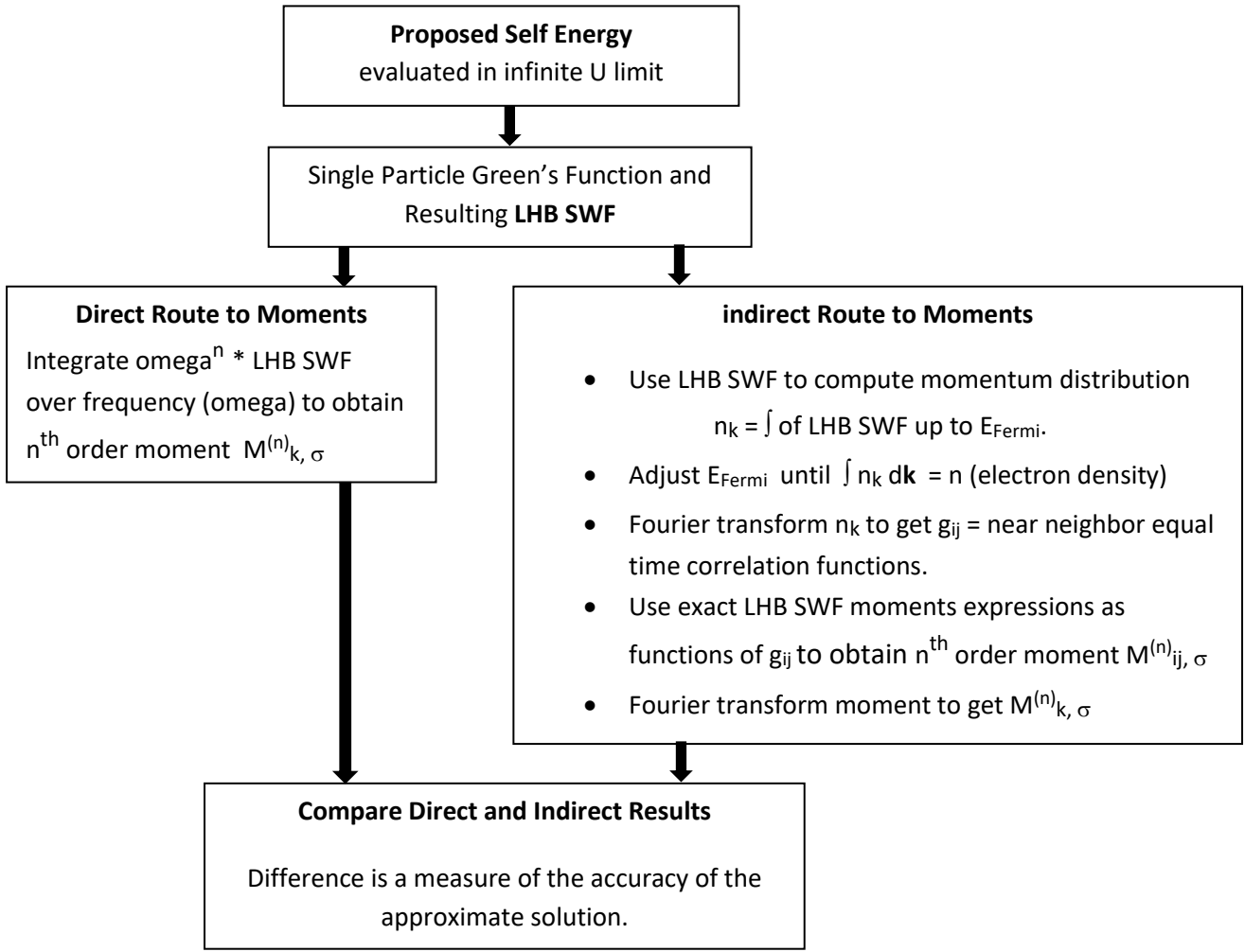

Figure 1. Flow diagram for comparing direct and indirect moment computations.

$$\Sigma(\omega) = -n\omega/(1-n) \qquad (1)$$

The question is whether this "strong coupling limit self energy" applies to the infinite U case. More accurately, the issue is whether the mean field hybridization function ([22], [31]) vanishes as U goes to infinity. Eqn. (8) of Park, et al. [32] indicates this is so for zero temperature within the DMFT approximation.[3] On the other hand, Jeschke and Kotliar [33] offer a DMFT solution invoking a decoupling approximation to the hybridization function and compares the resulting density of states to the density of states for a 6 atom cluster, the latter corresponding to a non-zero hybridization function consisting of five poles.[4]

---

[3] Parks et al. [32] maintain with this same equation that the Hubbard model kinetic energy vanishes at zero temperature, infinite coupling. This is yet another example of conflating the high temperature ("AL0") atomic limit with the physically interesting ("AL1" and "AL2") atomic limits where the off-diagonal equal time correlation functions are *not* zero when $\Delta \ll U$. Even for $\Delta \to 0$, the correlation functions are not zero.

For now, the self energy expression in Eqn. [1] will be taken at face value as representing the DMFT solution when $\Delta \ll U$ and will be designated as a "DMFT" solution. The Computer Lab tests whether even this simple, local self energy yields highly non-local (Indirectly computed) moments, with implications for the SWF width or scattering. The end of this Section will consider more general DMFT self energy expressions with non-zero hybridization functions.

Eqn. [1] leads to a single particle Green's function as

$$G(k, \omega) = (1-n)/(\omega - (1-n)\varepsilon_k) \quad (2)$$

where $\varepsilon_k$ is the Bloch energy. is the Bloch energy.

In turn, this yields a (LHB) SWF as

$$A(k, \omega) = (1-n)\delta(\omega - (1-n)\varepsilon_k) \quad (3)$$

The left hand path in Fig. 1 is trivial. The zeroth moment is $(1-n)$. The first moment is $(1-n)^2 \varepsilon_k$ so the energy dispersion – the ratio of the first and zeroth moment is

$$E(k, \omega) = (1-n)\varepsilon_k \quad (4)$$

The SWF width is the square root of the second central moment or

$$\sigma_k = \text{sqrt}(M^{(2)}(k)/M^{(0)} - E(k, \omega)^2) \quad (5)$$

Using the direct moment calculation with $A(k, \omega)$ as in Eqn. (3), the SWF width is zero as expected.

---

[4] While the agreement of the DMFT plus decoupling approximation to a finite cluster of atoms is interesting, the real test is how well the proposed solution behaves in the thermodynamic limit, to wit the Computer Lab of Fig. 1.

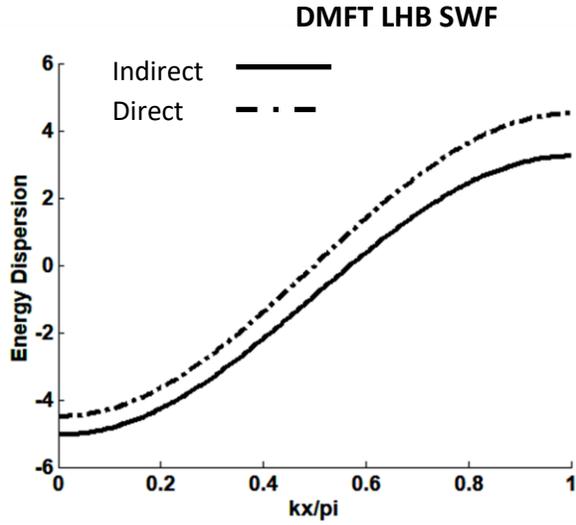

Figure 2. Comparison of energy dispersion computed by Direct (dash-dot) and Indirect (solid) method. Simple cubic lattice. DMFT self energy. n = 0.25

The more interesting results follow from the indirect method for moment calculations[5]. The delta function SWF in Eqn. (4) leads to a simple step function for the momentum distribution ( (1-n) for below the Fermi level and zero for above the Fermi level). With this and $_{ij}$using the 1D linear chain, 2D square lattice and 3D simple cubic nearest neighbor tight binding model for the Bloch energy (unit lattice spacing)

$$\varepsilon_k = (-1) \sum_1^D \cos(k_i) \quad i = 1...D \tag{6}$$

the Fermi energy and the near neighbor single particle equal time correlation functions $g_{ij}$ are easily computed (numerically). The values for $g_{ij}$ are inserted into the moment expressions derived in [13] and [3]. Eqns. (10-12, 25-27) in [3] already lay out all of the needed expressions for the first moment and resulting energy dispersion. The resulting energy dispersion is in Fig. 2, along with the direct moment energy dispersion of Eqn. (5). As is clear, the energy dispersions (i.e. first moments) are almost the same, aside from a mostly constant shift. That shift is mainly the "average kinetic" in Eqn. 11 of [3].

This shift, at least for the paramagnetic case, primarily just corresponds to a shift in the Fermi energy. However, as pointed out some time ago by Harris and Lange [12], the shifts are not equal for up and down electrons when the respective electron densities are different and has an important impact on, for example, ferromagnetic stability. In fact, as noted in [12], this shift is essential to correctly capture the well-known exact result of Nagaoka [35] that the ground state of the N site, N-1 electron infinite U Hubbard model is ferromagnetic.

---

[5] The first moment calculation is straightforward as presented in [3]. The second moment calculation is long and cumbersome. This paper relies on a second moment expression generated automatically using the very helpful Mathematica program "DiracQ" from Wright and Shastry [34] rather than the less reliable manual method used in [3]. The DiracQ second moment result is in the Appendix of [13]. The Mathematica program along with other technical details including the explicit expressions for the moments in terms of the single particle equal time correlation functions may be obtained by contacting the Author.

As expected, the comparison of the Direct and Indirect results for the widths of the SWF in the DMFT solution is much more substantial and interesting.

Fig. 3 shows the LHB SWF (Indirect moment) width $\sigma_k$ as computed with the Indirect method within the DMFT approximation (in this case, the infinite U SWF is determined by the local atomic limit self-energy of Eqn. (1) ). The SWF widths computed via the Indirect method show some similarity to the Single Peak SWF widths computed with a self-consistent SWF assuming a single Gaussian peak in Fig. 6. As noted, the DMFT LHB SWF is a single delta function, so the Direct method yields a zero width across all momentum  The local self-energy and (infinite U) delta function SWF are clearly at odds with the exact moment results computed via the Indirect method.

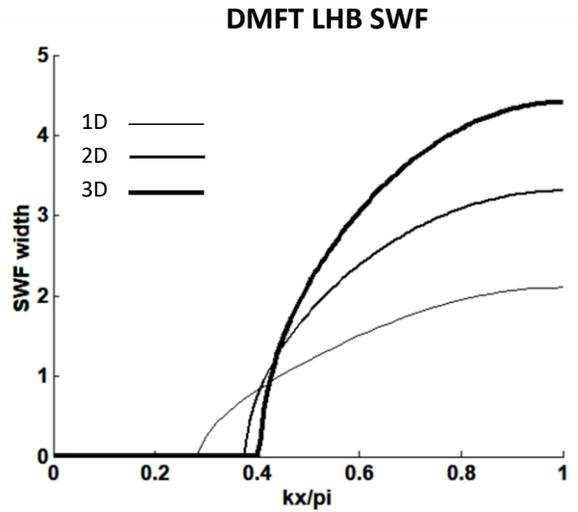

Figure 3. DMFT LHB SWF width $\sigma_k$ for n = 0.25

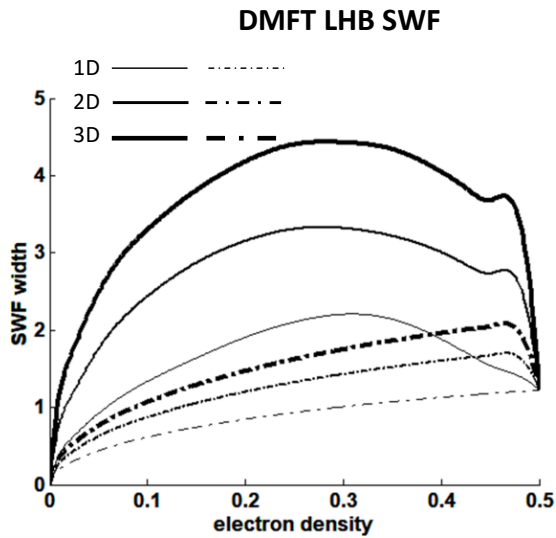

Figure 4. DMFT LHB SWF width as a function of electron density. Solid lines are $\sigma_{k,max}$ and dash-dot lines are $\sigma_{Zero}$, where each are defined in the text.

Fig. 4 reinforces the dichotomy between the Direct and Indirect SWF results. $\sigma_{k,max}$ is the maximum value of (Indirect Moment) $\sigma_k$ for a specific electron density. The Direct Method SWF width and $\sigma_{k,max}$ remain zero across all electron densities and lattice dimensions. Similarities with the Single Peak SWF results below (Fig. 7) are noted.

For reference, Fig. 4 also shows $\sigma_{Zero}$ as a function of the electron density. $\sigma_{Zero}$ is the SWF width evaluated using the Indirect method by taking all inter-site correlation functions $<c_{i\sigma}^\dagger c_{j\sigma}>$ as zero. $\sigma_{Zero}$ depends on the electron density and lattice structure/dimension but is independent of momentum (See Eqn. (7) below).

The deviation of $\sigma_{k,max}$ from $\sigma_{Zero}$ is yet another strong indication of the non-local nature of the SWF (and the underlying self-energy).

In a later Section, the LHB SWF widths are computed in a self-consistent variant of Fig. 1. The resulting widths are very similar to the DMFT (Indirect moment) widths (Compare Fig. 3 and 4 with Fig. 6 and 7).

The LHB SWF width (or second central moment), using the method of the companion paper [13], is expressed in terms of the electron density (n) and a limited number of near neighbor equal time correlation functions ($g_{ij}$). A systematic study of the SWF widths shows that they are moderately insensitive to the exact values of n and $g_{ij}$ within physically reasonable limits (n > g1 > g2 … > 0). This indicates that, whatever the exact moments might be, the exact widths of the LHB SWF will be similar to the behavior if Figs. 3, 4, 6 and 7. That is, the widths will show a strong momentum dependence (non-locality).

A central assumption of DMFT is the self energy $\Sigma$ is local [22, 31]. This means the imaginary part of $\Sigma$ is local or, equivalently, independent of momentum. Since the imaginary part of $\Sigma$ is closely linked to electron scattering and the width of the SWF [25], the conclusion is that any DMFT self energy expression, including non-zero hybridization functions, will necessarily lead to inconsistent Direct and Indirect Moments in the Computer Lab test of Fig. 1.

## Exact Results as a Computer Lab: Cluster Solutions

The restriction of the DMFT to a local self-energy (equivalently, no momentum dependence) has been a concern for some time. The single site DMFT solution has been extended to multi-site cluster solutions including the Dynamic Cluster Approximation (DCA) [31]. The DCA has a "weakly k-dependent self-energy" [31]. There do not appear to be any explicit expressions available for the self-energy or SWF for these cluster solutions specific to the strong coupling (infinite U) limit and general electron densities. However, Maier et al. [31] states that ".. the DCA become(s) exact in … the strong-coupling limit …(where) all the sites in the lattice are decoupled. The effective cluster problem reduces to a single-site problem without coupling to a mean field."

The statement that DCA – and other cluster mean field theories – become exact in the strong-coupling limit is based on a lack of appreciation of the subtleties of the Hubbard model atomic limit [13]. The DCA, along with many other solutions, is indeed exact in the less interesting high temperature limit (hopping much less than $k_B$ T) or **AL0** in [13]. But making this claim in the physically interesting low temperature limit ($k_B$ T much less than hopping and the latter much less than the intra-atomic potential U) is equivalent to claiming an exact solution for both the **AL1** ($k_B$ T = 0; hopping $\rightarrow$ 0; U finite) and **AL2** ($k_B$ T = 0; hopping finite; U $\rightarrow$ infinity) atomic limits of [13]. As explained decades ago [3, 12, 23], the physically interesting atomic limits (**AL1** and **AL2**) are non-trivial and not local. Put differently, as noted by Esterling [23], the claim to have an "exact solution in the strong-coupling limit" is equivalent to claiming an exact solution to a dynamic excluded volume problem (**AL2**) and, perforce, must yield identical results for the Direct and Indirect Moments calculated as in Fig. 1.

We hope this puts to rest the claim for cluster solutions (or indeed any solution) as becoming "exact" in the limit of $k_B$ T much less than hopping and hopping much less than U.

If in fact the DCA reduces to the DMFT solution in the "strong-coupling limit", then the above critique of the DMFT also applies to the DCA cluster solution. That is, these Computer Lab results call into question the reliability of the mean field theory LHB SWF centroids (quasi-particle energies) and widths (corresponding lifetimes) in the strong coupling limit and, by implication, in the strong coupling region (large but finite U).

On the other hand, if the mean field cluster solutions are not isomorphic with DMFT in the infinite U limit (for cluster sizes large than one atom), then the question is how well mean field cluster solutions capture the physics (quasi-particle energies and lifetimes) exposed with the exact moment results – in particular as regarding how well the solutions behave for limited cluster sizes.

The actual solution of the mean field equations for finite sized clusters, particularly at low or zero temperature, can be challenging [31]. However, extracting the moments of a cluster solution for the infinite U Hubbard case (the Hamiltonian and Green's functions in Eqns. (4) and (5) of [13]) does not need a complete solution. The moments can be computed from a high frequency expansion of the Green's function or equivalently a high frequency expansion of the self-energy, which follows directly from the definition of the Green's function in terms of the SWF [25]. As was pointed out by Bari and Lange [36] and by Esterling [23], in general attempts that seek a perturbative solution for the Hubbard model in powers of the hopping run afoul by neglecting terms of (hopping/omega) where omega is a characteristic frequency. Since the latter is on the order of the hopping, the series expansion does not converge and, indeed, can lead to contradictions [23]. But for high frequencies (much larger than the hopping), a series expansion is viable. Cluster solutions, including DCA, could extract numerical results for the low order moments of their solution – as a function of the cluster size – by expanding the cluster self-consistent equations in inverse powers of frequency, thus avoiding the complexity of cluster-specific methods.

This process can subject the cluster solutions to the proposed Computer Lab. The moments generated from the high frequency expansion will be the "Direct Moments" of Fig. 1. The "Indirect Moments" of Fig. 1 require explicit results for the single particle Green's function. But even if these are not known, good estimates of the Indirect Moments are available. As shown above for DMFT and, in the next Section for a single peak LHB SWF, the Indirect moments are relatively insensitive to the explicit solution. And in fact as the electron density approaches one electron per site, the exact results all converge to a non-trivial and large LHB SWF second central moment or width.

## Single Peak LHB Spectral Weight Function

Aside from the one dimensional case, virtually all solutions to the Hubbard model in the strong coupling region result in a single main peak in the lower Hubbard band (LHB) spectral weight function (SWF). The exact moment relations in the companion paper [13] can be used to generate a "best" (i.e. moment-preserving) single peak LHB SWF. It bears re-iterating that, while others have proposed similar moment-preserving SWFs, all previous solutions contain approximations to the required higher order correlation functions [14-17]. Our solution is unique in that the requisite equal time higher order correlation functions are *exactly* expressed as functions of the equal time single particle correlation function. By assuming a simple functional form (here a Gaussian) for the SWF, we determine the single particle correlation functions self-consistently from the SWF as detailed below.

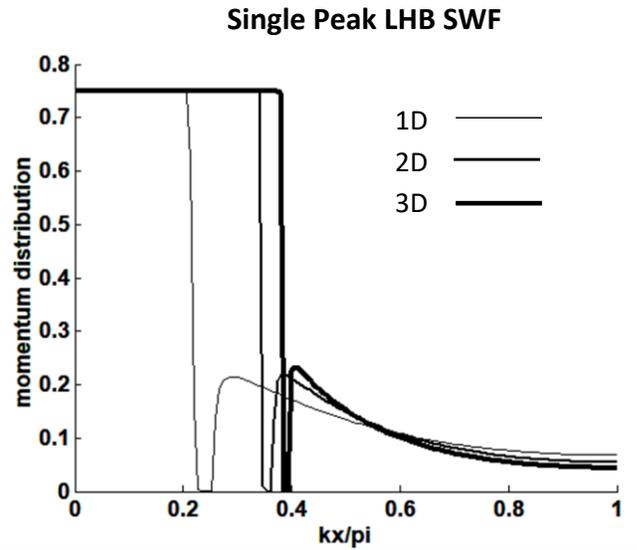

Figure 5. Momentum distribution for n = 0.25

Revisions and implications for the single peak LHB SWF assumption are taken up later. Some numerical results are sensitive to this assumption, others not so. At the conclusion of this Section, this distinction will be made. To provide contrast and comparison with the DMFT (local self-energy) solution of the previous Section, emphasis will be on the implications of the solution for strong non-locality effects in the SWF and the corresponding self-energy.

Expressing the zeroth, first and second moments as functions of the single particle equal time correlation function $<c^\dagger_{i\sigma} c_{j\sigma}>$ as in [3] and the Appendix of [13] and assuming a Gaussian functional form for the SWF, we determine a self-consistent solution for $<c^\dagger_{i\sigma} c_{j\sigma}>$ and the SWF for a given electron density n. (The Fermi energy is determined as well in the usual way as in [3] (Eqns. 31 and 32) ).

The self-consistent strategy mimics the Indirect method of Fig. 1, but the process is to start, for a given electron density n, with an initial guess for the near neighbor correlation functions ($g_{ij}$), loop through the computations ending up with output near neighbor correlation functions ($g_{ij}$). Compare input with output and adjust the Fermi energy and $g_{ij}$ until consistency.

As in the preceding Section, the results are for a 1D linear chain, a 2D square array and a 3D simple cubic lattice with nearest neighbor hopping, finite Δ, infinite U and unit lattice spacing. The method and formulae are easily extended to higher dimensions and alterative lattice structures. Focus is on the momentum distribution (see ([3] Eqn. 31) ) and the SWF width, computed as in ([3] Eqn. 30).

Fig. 5 shows the typical strong momentum-dependence (non-locality) of the momentum distribution (the Fourier Transform of $<c^\dagger_{i\sigma} c_{j\sigma}>$ for the 1D, 2D and 3D lattices with an electron density n = 0.25 (Recall n is half the total electron density for the paramagnetic case). The momentum direction is kx for 1D, **k** = (kx, kx) for 2D and **k** = (kx ,kx, kx) or body diagonal for 3D.

The odd valley in the momentum distribution for the 1D linear chain is purely an artifact of the 1D SWF width going to zero for certain kx/pi values. There are three regions (see Fig. 6). The first has the SWF width as zero with the momentum below the Fermi momentum. In this region, the SWF is a delta function and the momentum assumes the full weight of the SWF (zeroth moment) integrated up to the Fermi energy leading to a value of (1-n) or 0.75 for Fig. 5. The second region has the SWF width still zero but the momentum is above the Fermi momentum so the integral of the delta function SWF up to the Fermi energy yields a zero value for the momentum distribution. In the third region, the SWF width is not zero. The computed SWF has a tail that extends below the Fermi momentum (Fermi energy) leading to a non-zero but decreasing momentum distribution.

As already noted, a non-constant momentum distribution over k necessarily implies that its Fourier transform $<c_{i\sigma}^\dagger c_{j\sigma}>$ is non-local.

The SWF width provides more interesting results.

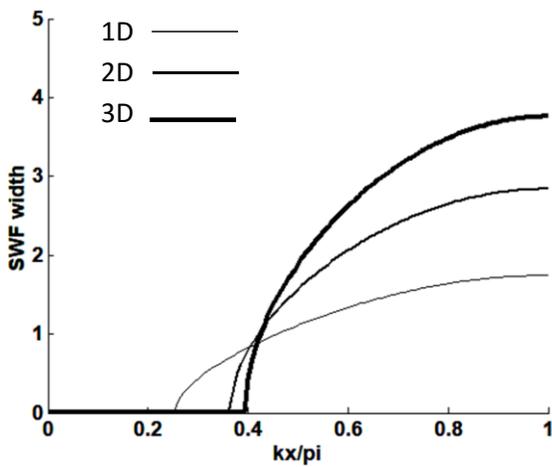

Figure 6. LHB SWF width $\sigma_k$ for n = 0.25

Fig. 6 shows the variation in the SWF width $\sigma_k$ again for n = 0.25 along the same **k** values as Fig. 5.

Once again, the non-local (k-dependent) nature of $\sigma_k$ and the SWF is apparent.

As a reminder, the Dynamic Mean Field Theory (DMFT) "strong coupling" solution [22] to the infinite U Hubbard model predicts a zero width to the LHB SWF across all momentum (and electron densities), all lattice structures and all dimensions.

Fig. 7 shows the maximum SWF width $\sigma_{k,max}$ as a function of electron density n, from n= 0 to n = 0.5 (one electron per site). The maximum SWF width, $\sigma_{k,max}$ is the largest width for a given electron density over the first Brillouin zone.

Fig. 7 also shows $\sigma_{Zero}$, defined earlier as part of Fig. 4, as a function of the electron density. $\sigma_{Zero}$ depends on the electron density and lattice structure/dimension but is independent of momentum as in Eqn. (7) below.

Once again, the deviation of $\sigma_{k,max}$ from $\sigma_{Zero}$ is another strong indication of the non-local nature of the SWF (and the underlying self-energy).

Anticipating the analysis when the single assumption of a single peak in the SWF is removed, there should be no doubt that a correct SWF width (more accurately, second central moment) and the corresponding self-energy must include a significant non-local component, even in the strong coupling region.

What is not expected, based on experience with weakly interacting electrons, is the width remaining large even as the density approaches one electron per site, so the hole density (empty sites) is small.

The source of the large $\sigma_k$ near one electron per site is easy enough to trace mathematically using the second moment of the lower peak SWF as given in the Appendix of the companion paper [13]. If the single particle equal time correlation function $< c_{i\sigma}^\dagger c_{j\sigma} >$ for different site indices is set to zero, then many of the terms simplify as in that Appendix. In fact, this is exact as the electron density approaches one electron per site.

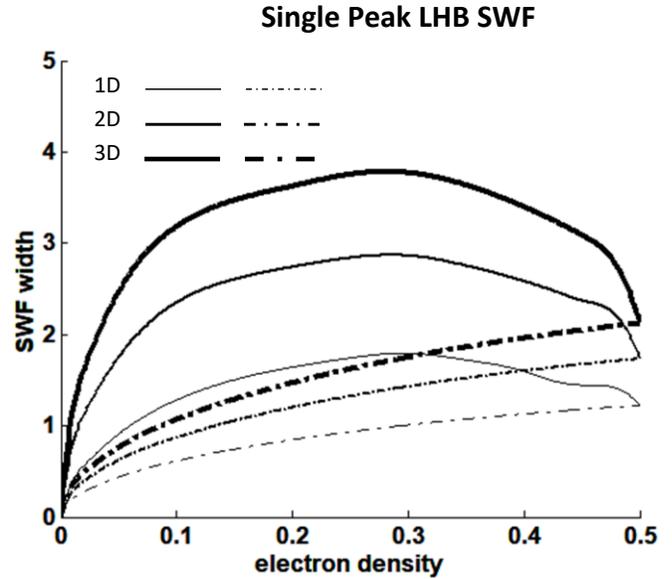

Figure 7. LHB SWF width as a function of electron density. Solid lines are $\sigma_{k,max}$ and dash-dot lines are $\sigma_{Zero}$, where each are defined in the text.

Using the exact localized correlation functions for one electron per site in the moment expressions in the Appendix of [13], all of the terms in the local (i=j) second moment vanish except the first term. Similarly examining all of the terms in the non-local i≠j second moment expression, the only non-zero terms involve certain density-density equal time correlation functions. In the localized limit, these are $<n_{i-\sigma} n_{j-\sigma}> = n_{-\sigma}^2$ for i≠j and $<n_{i-\sigma} n_{p-\sigma} n_{j-\sigma}> = n_{-\sigma}^3$ for i≠p≠j. So the second moment M2$_{ij\sigma}$ for general i, j in the Appendix becomes $\Delta^2$ *t$_{ip}$ t$_{pi}$ *(1- $n_{-\sigma}$)$^3$ + $\delta_{ij}\Delta^2$ * t$_{ip}$ t$_{pi}$ *[ (1- $n_{-\sigma}$) - (1- $n_{-\sigma}$)$^3$ ]. Taking the Fourier transform of M2$_{ij\sigma}$ and calculating the second central moment as defined in ([3], Eqn. 30) ) with the corresponding first moment for one electron per site or (1- $n_{-\sigma}$)$^2$ $\Delta\varepsilon_k$ , leads to a k-independent SWF width which is the same as $\sigma_{Zero}$ referenced above.

$$\sigma_{k,\sigma} = \sigma_{Zero} = \Delta\,(Z\,n_{-\sigma}\,(2-n_{-\sigma}))^{1/2} \qquad (7)$$

where Z is the number of nearest neighbors.

This result for the second central moment for the LHB SWF for infinite U, finite $\Delta$ and zero temperature as the electron density approaches one electron per site *follows simply and directly from the definition*

*of the SWF.* There are no further assumptions. This is a strong constraint on any proposed solution to the Hubbard model in the strong coupling region.

The results are based on a rigorous solution of the Hubbard model in the strong coupling limit. Up to this point, the assumption of a Gaussian form for the lower peak SWF is the *only* assumption in this Section, other than working in the infinite U, finite $\Delta$ limit. The non-local effects are so strong that the general conclusions of this paper should not be impacted if an alternative 3 parameter form (e.g. parabolic or Lorentzian) were assumed or if higher order moments such as skewness were included. A non-local self-energy with a substantial non-local imaginary part even for one electron per site must be included in any plausible solution to the Hubbard model.

There are alternative expressions for the SWF where the width (or scattering if considered as proportional to the SWF width) goes to zero as the electron density approaches one electron per site. The simplest functional extension is a two (or more) peak structure as, for example, in one dimensional solutions to the Hubbard model [37-41]. A simple case is two peaks symmetrically situated and structured relative to the first central moment or Ek as given in Eqns. (10a-10b and 29) of [3]. Finding the actual (k and n-dependent) weights and widths of the two sub-peaks will entail determining the third, fourth and fifth central moment which is beyond the scope of this specific computation. However, if for now we make the simple assumption of modeling the low energy structure of the SWF by two delta functions with equal weights, the moments are replicated if the delta functions are located at energies Ek + $\sigma_k$ and Ek - $\sigma_k$ each with a weight of (1-n)/2 and width $\sigma_k$ derived from the second central moment as in Eqn. (5). There will be some variance in $\sigma_k$ from its single Gaussian peak value since the momentum distribution and resulting equal time single particle correlation functions will be a bit different.

Certain quantities are sensitive to the choice of the specific SWF functional form. For example, the momentum distribution can change dramatically for multiple peaks. The current "smeared out" form in Fig. 2 can be traced to the use of a single Gaussian peak. But multiple peaks will only sharpen that structure, leading to a larger – not smaller – non-local single particle correlation function $< c_{i\sigma}^\dagger c_{j\sigma} >$ as the Fourier transform of the momentum distribution. The total energy depends directly on the momentum distribution [36]. Hence numerical computations of the total energy and resulting stability issues such as ferromagnetic versus paramagnetic states relying on a solution derived from or relying solely on moments must be viewed with caution.

On the other hand, as already noted, the width of the SWF (second central moment) is a function of the electron density (n) and the nearest neighbor single particle correlation functions (the g$_{ij}$ in Fig. 1). Our experience with the self-consistency computations described above indicates that as long as the electron density n > g1 > g2… > 0 where g1, g2 are the progressive near neighbor single particle correlation functions, the first and second moments are moderately insensitive to the actual numerical values. Compare, for example, Fig. 3 with Fig. 6 and Fig. 4 with Fig. 7, using very different assumptions on the infinite U SWF. The significant non-local nature of the SWF width will be retained for any reasonable values for these parameters.

## Direct versus Indirect Moments

A reasonable question is whether the moment results using the Direct or the Indirect methods of Fig. 1 are more trustworthy. The Direct Method moments speak directly to the physics, right or wrong, in the approximate solution and LHB SWF. The Indirect method, we will now argue, offers a more accurate and physically correct estimate of the moments. So the deviation of the Direct method moments from the Indirect moments (or from the Single Peak moments of the preceding Section) is a simple but significant test of the quantitative and even qualitative fidelity of the approximate solution.

Consider the first moment of the LHB SWF or, more particularly, the corresponding energy dispersion. That energy dispersion is exactly given by Harris and Lange [12] as

$$E(k) = (1-n)\, \varepsilon_k - \tau/(1-n) + L_k/(1-n) \qquad (8)$$

where $\tau$ (the average kinetic energy) and $L_k$ are defined in Eqn. 11 & 12 of Esterling and Dubin [3]. We reiterate this expression is exact. $L_k$ is the Fourier transform of two-particle equal time correlation functions. As Esterling and Dubin point out, $L_k$ scales as the square of g, where g is a near neighbor single particle correlation function. On the other hand, the first and second terms in Eqn. 9 scale as n and/or the first power of g. So a reasonably accurate estimate of the energy dispersion has a band-narrowing term and a k-independent (average kinetic energy) shift:

$$E(k) = (1-n)\, \varepsilon_k - \tau/(1-n) \qquad (9)$$

This is just what is shown in Fig. 3 (DMFT Indirect Method case). The Direct method energy dispersion has the band-narrowing term, but entirely misses the k-independent energy shift. As already noted, this energy shift plays an important role in ferromagnetic stability so contains important physics.

The Single Peak LHB SWF in Fig. 6 and 7 follows Eqn. 8, estimating $L_k$ self-consistently rather than setting it to zero.

The physics of the LHB SWF width or, more precisely $\sigma_k$ as derived from the second central moment in Eqn. (5) is much more complex as the expression (see the Appendix of [13] ) is more complicated. But the comparison simplifies as the electron density approaches one electron per site (n approaches 0.5). In that limit, all the sites become occupied and – for infinite U – no inter-site hopping can occur. So the equal time correlation functions are local and trivial. In that region $\sigma_k$ becomes independent of momentum k and is $\sigma_{Zero} = \Delta\,(Z\,n\,(2-n)\,)^{1/2}$ where Z is the number of nearest neighbors. This is an exact result. There are no estimates of the equal time correlation functions since they exactly vanish or are trivial (a power of n). The value of the second central moment and $\sigma_k$ is certain in that limit. What is unresolved is whether this $\sigma_k$ is a measure of a single peak SWF or whether the result points to a LHB SWF with multiple peaks.

There is a clear contrast with the Direct Method moments. For the DMFT approximation, using the "strong coupling" self energy of Eqn. {1}, the Direct Moment $\sigma_k$ is zero in this limit, a violation of an exact and physically important result. This exact result for the second central moment of the LHB SWF as the electron density approaches one electron per site (Eqn. (7)) will be a severe test for any approximate solution, including cluster mean field theories, to the Hubbard model in the strong coupling region.

## Infinite Dimensions

The DMFT and cluster mean field extensions are asserted to be exact in infinite dimensions or coordination number [19-22, 31]. To avoid certain expressions from being unbounded, $\Delta$ is assumed to scale as $Z^{-1/2}$ (equivalently, for large dimensions D, as $D^{-1/2}$ ). Using Eqn. (7) above, this means that $\sigma_k$ near one electron per site becomes $\Delta (Z\, n\, (2-n)\,)^{1/2}$ which is a constant, not zero. The necessary conclusion is, for this aspect of the solution, DMFT and – likely – cluster mean field extensions which predict exactly zero second central moments for strong coupling are not exact. However, the bandwidth scales as $\Delta Z$, so the ratio of $\sigma_k$ to the bandwidth scales as $Z^{-1/2}$ or $D^{-1/2}$ goes to zero as the dimensions and/or coordination number grows to infinity.

## One Dimension

The one dimensional Hubbard model has an exact solution [11]. But this solution is for the ground state energy. Properties of excited states, correlation functions and the SWF rely on some fairly sophisticated computations [37-41]. The SWF, including the infinite U limit, has been shown to have at least two main peaks ("holons" and "spinons") as well as a third weaker ("shadow") peak [40]. Our moment results are silent on the explicit functional form for the SWF, including the number and shape of dominant peaks.

But the moment relations do offer an assessment of the one dimensional computations that seek to extend the exact 1D results.

Recently Nocera, Essler and Feiguin [41] have conveniently offered an explicit expression for the 1D Hubbard model SWF in the infinite U limit and one electron per site (n = 0.5). Using

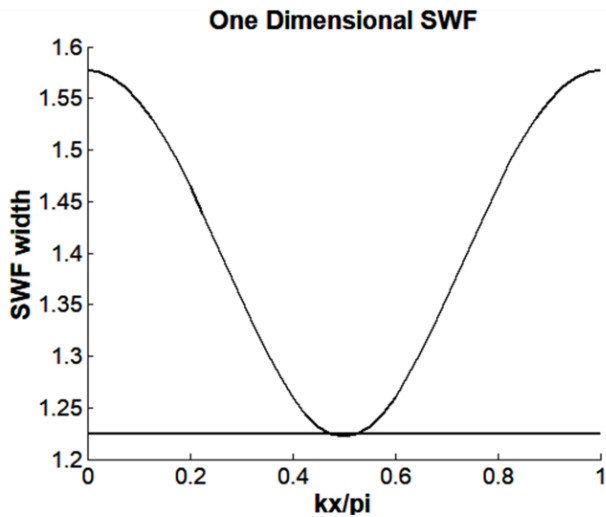

Figure 8. SWF width for one electron per site and infinite U for the one dimensional Hubbard model using Nocera, Essler and Feiguin [41]. The solid line at the base is the exact result 1.2247 $\Delta$

that SWF in the Indirect moment computations, the zeroth moment is 1-n or 0.5 as expected. But the first moment is virtually zero over all momentum values, leading to an energy dispersion of zero rather than the expected (local equal time correlation functions result)

$$E(k) = (1-n) \varepsilon_k = 0.5 \varepsilon_k \qquad (10)$$

The SWF shows two well separated peaks, with the second central moment k-dependent as in Fig. 8. This contradicts the expected k-independent second central moment width of $\Delta (Z n (2-n))^{1/2}$ = 1.2247 $\Delta$ (solid line at the bottom) for Z = 2 nearest neighbors for a one dimensional chain for one electron per site.

The zero energy dispersion (momentum independence) for the first moment with the Nocera, Essler and Feiguin [41]solution indicates a local first moment, in contrast to the result of Eqn.(10). For one electron per site and infinite U, the electrons are trapped in place. Where does the dispersion come from?

A simple way to answer this question is to consider a similar problem: a spinless Hubbard model with U set to zero. The solution is trivial for all densities. Consider the electron density approaching or at one electron per site. Once again the electrons are trapped. But the first moment is $\varepsilon_k$, not zero. The explicit expression for the SWF as in Eqn.(2.8) of Harris and Lange [12], taken at zero temperature may help. Using their notation, the "a" state is just a set of free electrons up to the Fermi energy. The "b" (N+1 electron) state – for non-zero matrix elements— has one electron above the filled Fermi sea. The "d" (N-1 electron) state – for non-zero matrix elements— has one electron missing in the filled Fermi sea. Putting in the usual expression for the Wannier operators as a sum of Bloch state operators, and then integrating the resulting real space SWF * frequency over frequency yields a first moment of $t_{ij}$, the Fourier transform of $\varepsilon_k$. This remains so even as the density approaches or equals one electron per site. In that case, only the second term in Eqn. (2.8) of [12] contributes.

These results for a nominally exact solution for the one dimensional Hubbard model suggests that each of the current 1D solutions, e.g. [37-40] would benefit from an moment analysis as in Fig. 1. The point here is that the exact moment tests are based on almost trivial mathematics, using only the definition of the SWF. On the other hand, the 1D solutions invoke quite sophisticated methods. Simplicity is always a desirable trait.

## Falikov Kimball Model

Falikov and Kimball [42] offer a model similar to the Hubbard model, but much simpler. For the infinite U limit, the Hamiltonian and single particle Green's functions are as in Eqns. (4) and (5) of [13], but there is no sum over spin ($\sigma$) in Eqn. (4). The down electrons have no hopping. The moments are quite easy to compute and are (using the notation in [13]), for a homogenous system, no special short or long range order and an implied sum over repeated indices):

$$M0_\sigma = (1 - n_{-\sigma}) \qquad (11)$$

$$M1_{ij\sigma} = \Delta\, t_{ij} < (1 - n_{i-\sigma})(1 - n_{j-\sigma}) > \qquad (12)$$

$$M2_{ij\sigma} = \Delta^2 * t_{ip} t_{pj} * < (1 - n_{i-\sigma})(1 - n_{p-\sigma})(1 - n_{j-\sigma}) > \qquad (13)$$

For one electron per site, the density-density correlation functions factor for different sites (are local) and, taking care of the i=j case so that $(1 - n_{i-\sigma})(1 - n_{j-\sigma})$ becomes $(1 - n_{i-\sigma})$ then

$$M2_{ij\sigma} = \Delta^2 * t_{ip} t_{pj} * [(1 - n_{-\sigma})^3 + \delta_{ij}\, n_{-\sigma}(1 - n_{-\sigma})^2 \qquad (14)$$

and the second central moment becomes $\Delta^2 Z\, n_{-\sigma}(1 - n_{-\sigma})$ so the width is

$$\sigma_{k,\sigma} = \sigma_{Zero} = \Delta\, [Z\, n_{-\sigma}(1 - n_{-\sigma})]^{1/2} \qquad (15)$$

Even for this very simple model, as the electron density approaches one electron per site there is a large non-zero second central moment for the SWF except of course for the trivial case of $n_{-\sigma} \to 1$.

## Conclusions

The moments of a function alone cannot, in general, uniquely determine the explicit function. However SWF moments calculated according to Fig. 1 offer a rigorous test of any solution to the Hubbard model. The proposed single particle Green's function can be inserted into the exact SWF moment expressions. The resulting moments can compared with the moments calculated from an explicit integral of the proposed SWF, weighted by the frequency to the appropriate power. Any difference between the two results provides a direct quantitative and qualitative test of the solution.

Any "single main low (and high) energy peak" solution for the SWF must comply with the large non-local widths found here. This is particularly true as the electron density approaches one electron per site where the numerical value of the width is independent of SWF functional form as the single and multi-particle correlation functions become local and trivial.

The Single Peak SWF derived here is based on the zeroth, first and second moments, computed self-consistently. An improved solution would include the third moment to take skewness into account. See, for example Fig. 4 in [12] where the improved Hubbard solution [2] shows considerable skewness for one electron per site, paramagnetic. However, yet higher order moments (fourth moment or kurtosis) are likely to offer diminishing returns on the predictive power of the Single Peak self-consistent moment solution.

Figs. 4 and 6 show that $\sigma_{Zero}$ is a significant component of the total SWF width and, perhaps, could be the basis for two or more peaks into the low (and high) energy SWF structure. An alternative solution, consistent with the results so far and adding yet another speculative Hubbard model solution, would be to assign the spacing between the two peaks in the low energy SWF structure to $\sigma_{Zero}$ with the width of the sub-peaks equal and determined by the difference between $\sigma_k$ and $\sigma_{Zero}$. These latter widths will go to zero for low density and for one electron per site. A better solution is to compute the third through fifth central moments of the SWF and use this data to fit the weights, separation and widths of a dual peak SWF.

The specific results in this paper are for a 1D linear chain, 2D square and 3D simple cubic lattice, but most of the relationships are valid for any dimension and lattice structure. Computation of the first three moment relations and ensuing results as a function of system dimensions will provide insight into when the "large dimension" DMFT and cluster solutions become numerically relevant. However, for physically interesting (one to three) dimensions, the assumption of a local self-energy by the DMFT clearly seems to lead to sizeable errors, particularly in the second central moment which is related to electron lifetimes. Lacking an explicit expression for the cluster LHB SWF for finite $\Delta$, infinite U, no specific computations have been done but, to the extent that the infinite U cluster results are similar to the DMFT results, a corresponding concern applies.

This is a mathematical presentation, generating rigorous and demanding benchmarks for any solution to the Hubbard model that purports to be accurate in the strong coupling region. The analysis here is based on extremely simple, if tedious, mathematics – in sharp contrast to the specialized techniques employed elsewhere, especially for 1D solutions. DMFT and its cluster extensions have been used to explain a number of experiments ([22, 31] and references therein). But, as noted earlier, the Hubbard model is an extreme simplification of actual material behavior. The model simplicity while still including physically significant effects may well drive some of the agreement with experiment. The real issue is whether any proposed solution captures the essential physics of the *model itself*, without conflating actual material complexity. This Computer Lab offers just such a rigorous, unbiased test that includes some basic physical quantities (quasiparticle energy and width or scattering) applicable to the important strong coupling region.

### Afterward – Lessons Learned and a Challenge

- Claims to be "exact in the atomic limit" must address the physically interesting limit of $k_BT \ll \Delta \ll U$, not the high temperature limit of $k_BT \gg \Delta$.
- At zero temperature, there are only two energy scales ($\Delta$ and U), so any claim to be exact for vanishingly small hopping ($\Delta$) is tantamount to claiming an exact solution for infinite U. ($\Delta = 10^{-6}$ and U = 1 is the same as $\Delta = 1$ and U = $10^6$, only a change in units). A general and exact solution for infinite U (aside from special cases such as one dimension) is unlikely.
- Again noting there are only two energy scales at zero temperature, then – as used in the companion paper [13] – equal time correlation functions are the same whether computed for vanishing hopping (**AL1**) or infinite U (**AL2**). This leads to a host of interesting yet exact relations among equal time correlation functions for infinite U [13].
- In view of the preceding, claims to be "exact in the atomic (or strong coupling) limit" such as asserted by mean field theory solutions [22, 31] are not tenable and cannot be used to justify proposed solutions for intermediate coupling.

- The second central moment (width) of the LHB SWF has a strong momentum-dependence and evolves to a large, exactly computed value at one electron per site.
- DMFT assumes a local self energy so cannot replicate the strong momentum dependence of the LHB SWF width or corresponding self energy.
- Claims to be exact in infinite dimensions is interesting, but the real world has few dimensions. The Computer Lab offers a path to test how rapidly any solution converges with increasing dimensions or coordination number.
- An open question is whether the "weakly k-dependent self energy" of cluster mean field solutions [31] – or indeed any of the many extant proposed solutions [6] – can pass the test offered by this Computer Lab.
- The exact 1D solutions for the SWF would benefit from the Computer Lab test. If the solutions are truly exact, the Direct and Indirect moments should be identical.

This is the challenge – and opportunity – offered by the Computer Lab.

## Acknowledgements

With pleasure, we acknowledge the inspiration of R.V. Lange. Numerous helpful conversations with B. S. Shastry are also gratefully acknowledged.